\documentclass[amsmath,amssymb,10pt, twocolumn]{revtex4}
\usepackage{hyperref, graphicx, amsmath, amsmath, amssymb}
\newcommand{\vphi}[0]{\delta\phi}

\newcommand{\half}[0]{\frac{1}{2}}

\newcommand{\ld}[0]{\mathcal{L}}

\newcommand{\defn}[0]{\equiv}

\newcommand{\qsubrm}[2]{{#1}_{\scriptsize{\textrm{#2}}}}

\newcommand{\sol}[0]{\ld_{\scriptscriptstyle\{2\}}}

\newcommand{\tis}[0]{ {\theta}^{\scriptscriptstyle\rm{S}}}
\newcommand{\tisdot}[0]{ {\dot{\theta}}^{\scriptscriptstyle\rm{S}}}
\newcommand{\pis}[0]{ {\Pi}^{\scriptscriptstyle\rm{S}}}
\newcommand{\piv}[0]{ {\Pi}^{\scriptscriptstyle\rm{V}}}
\newcommand{\pit}[0]{ {\Pi}^{\scriptscriptstyle\rm{T}}}

\newcommand{\kin}[0]{{\mathcal{X}}}
\newcommand{\hct}[0]{\mathcal{H}}
\renewcommand{\figurename}{Figure}
\newcommand{\ep}[0]{{ {\delta}_{\scriptscriptstyle{\rm{E}}}}}
\newcommand{\lp}[0]{{ {\delta}_{\scriptscriptstyle{\rm{L}}}}}
\def\be{\begin{equation}}
\def\ee{\end{equation}}
\def\bea{\begin{eqnarray}}
\def\eea{\end{eqnarray}}
\def\bse{\begin{subequations}}
\def\ese{\end{subequations}}
\newcommand{\lied}[1]{\pounds_{#1}}

\let\oldsqrt\sqrt
\def\sqrt{\mathpalette\DHLhksqrt}
\def\DHLhksqrt#1#2{%
\setbox0=\hbox{$#1\oldsqrt{#2\,}$}\dimen0=\ht0
\advance\dimen0-0.2\ht0
\setbox2=\hbox{\vrule height\ht0 depth -\dimen0}%
{\box0\lower0.4pt\box2}}

\begin{document}
\title{Parameterizing dark sector perturbations  via equations of state}
\author{Richard A. Battye}
\email{richard.battye@manchester.ac.uk}
\affiliation{Jodrell Bank Centre for Astrophysics, School of Physics and Astronomy, The University of Manchester, Manchester, M13 9PL, U.K.}
\author{Jonathan A. Pearson}
\email{jonathan.pearson@durham.ac.uk}
\affiliation{Centre for Particle Theory, Department of Mathematical Sciences, Durham University, South Road, Durham, DH1 3LE, U.K.}
\date{\today}
\begin{abstract}
The evolution of perturbations is a crucial part of the phenomenology of the dark sector cosmology. We advocate parameterizing these perturbations using equations of state for the entropy perturbation and the anisotropic stress. For small perturbations, these equations of state will be linear in the density, velocity and metric perturbations, and in principle these can be related back to the field content of the underlying model allowing for confrontation with observations. We illustrate our point by constructing gauge invariant entropy perturbations for theories where the dark sector Lagrangian is a general function of a scalar field, its first and second derivatives, and the metric and its first derivative, ${\cal L}={\cal L}(\phi,\partial_\mu\phi,\partial_\mu\partial_\nu\phi,g_{\mu\nu},\partial_{\alpha}g_{\mu\nu})$. As an example, we show how to apply this approach to the case of models of Kinetic Gravity Braiding.
\end{abstract}

\maketitle

\textit{\textbf{Introduction}} The last decade in cosmology has seen the detection and measurement of a large number of observables which are sensitive to the details of the dark sector and its perturbations \cite{2012arXiv1201.2434W}. This has received a boost recently with the release of the results from the \textit{Planck} satellite \cite{Ade:2013xsa, Planck:2013kta}, and future missions such as  \textit{Euclid} \cite{Laureijs:2011mu, Amendola:2012ys} and \textit{COrE} \cite{2011arXiv1102.2181T} will make this area one of the most exciting in the coming decade. A consistent model of the gravitational dynamics of the dark sector is essential if one wants to meaningfully use  data to tell us about which of the plethora of dark energy/modified gravity theories \cite{Copeland:2006wr, Durrer:2008in, Clifton:2011jh, dakrenergy_amendola} could be realized by nature. Both the behavior of the equation of state parameter $w=P/(\rho c^2)$ and perturbations in the dark sector must be modeled; neglecting dynamics of perturbations of the dark sector is only valid when $w=-1$.  

For a given Lagrangian of the dark sector, the equation of state $w(a)$ and the perturbation evolution can be deduced from the Euler-Lagrange equations. However, at present there is no compelling fundamental model and so many models have been proposed that some kind of phenomenological approach would appear to be in order that could act as a staging post between observations and fundamental theories. Of course, only one could possibly be the true model and many of the models proposed are not entirely consistent as fundamental theories, but taking advantage of the projected observational advances must be part of levelling in on the true theory. Developing such an approach is the main objective of this \textit{letter}.

We will presume that the dark sector can be modelled by an effective energy-momentum tensor $U_{\mu\nu}$ which is compatible with the symmetry of the Friedmann-Robertson-Walker solution to Einstein's equations. Such a model can be entirely defined by the contribution of dark sector at the level of the background by specifying $w(a)$ and indeed most papers presenting observational results specify constraints on $w$, albeit at present usually assuming that it is constant since the data are not currently particularly constraining. Recent results from a principal component analysis study \cite{Zhao:2012aw} suggest that  model and parameterization   independent constraints on $w$ will be possible in the future. Moreover, under some weak assumptions, in some of the simplest models for the dark sector, for example, minimally coupled Quintessence models specified by a potential $V(\phi)$ or $F(R)$ modified gravity models, there is a one-to-one correspondence between the free function(s) describing the model and $w(a)$. Therefore, we consider this aspect of the phenomenology of the dark sector to be under control, if not completely solved.

The situation is less clear for the evolution of perturbations. One approach that has been proposed is to parameterize our ignorance of the dark sector into functions of time and space that  can be interpreted as a modified gravitational constant and gravitational ``slip" \cite{PhysRevD.81.083534,PhysRevD.76.023507,PhysRevD.77.103524,Silvestri:2013ne}.  A complimentary approach, which is much closer to that advocated here, has been developed in  \cite{Skordis:2008vt, Baker:2011jy, Zuntz:2011aq, Baker:2011wt, Baker:2012zs}. These works have developed model independent modifications to the perturbed Einstein equations which are gauge invariant. In both cases the general nature of these approaches, perceived as a virtue in their development, could also be a weakness since this leads to large number of free functions of space and time making it difficult to come significant conclusions about the nature of the dark sector.

Our ``holy grail'' is to provide a parameterization which assumes almost zero phenomenological prejudices, and all freedom is able to be traced back to some underlying symmetry or physically motivated principle. For instance, one could model the breaking of Lorentz or reparameterization invariance and then devise observations to constrain the magnitude of the violating terms. Our approach will still have a number of free functions, but these will be just of time and not space, which is a significant simplification. They will be minimal for a given set of assumptions and importantly it will be possible to connect them to general aspects of the underlying physical model. 

We will build on our earlier work \cite{PhysRevD.76.023005,Battye:2012eu, BattyePearson_connections} which could be described as a background fixed effective field theory approach \cite{Creminelli:2008wc,Gubitosi:2012hu, 2012arXiv1211.7054B} modelling those degrees of freedom that are observationally active  -- not to be confused with the background dependent approach, for example \cite{Bloomfield:2011wa}. The active degrees of freedom we consider here are scalar fields, the metric and their derivatives. In addition, one could include vector fields but we will not consider that possibility here. We will impose reparametrization invariance and second order field equations. An important novel feature of the approach we advocate is then to eliminate the active degrees of freedom to establish an equation of state for the perturbations that will be linear in the density, velocity and metric perturbations. What we are doing is modelling the imperfect fluid behaviour, that is due to the complicated internal degrees of freedom, with this equation of state and this is, of course, what we also do at background order. A restricted case of this the general idea was argued for in \cite{Sawicki:2012re} without the specific connection to the internal degrees of freedom.

\textit{\textbf{Basic formalism}} The perturbed field equations are $\ep {G^{\mu}}_{\nu} = 8 \pi G \ep {T^{\mu}}_{\nu} + \ep {U^{\mu}}_{\nu}$ where $\ep$ denotes an Eulerian variation.  The  perturbed dark energy momentum tensor $\ep {U^{\mu}}_{\nu}$  contains all information about the dark sector at the level of linearized perturbations, and its components    can be identified with perturbed fluid variables,
\bea
\label{eq:sec:ep_u_flui-intro}
\ep {U^{\mu}}_{\nu} = \delta\rho u^{\mu}u_{\nu} + 2 (\rho+P)v^{(\mu}u_{\nu)} + \delta P {\gamma^{\mu}}_{\nu} + P{\Pi^{\mu}}_{\nu}. 
\eea
The perturbed fluid variables $\{\delta\rho, v^{\mu}, \delta P, {\Pi^{\mu}}_\nu\}$  are constructed from perturbed field variables which appear in the underlying theory in a   complicated way (we will give an explicit example later). One can perform the standard split of the anisotropic stress, ${\Pi^{\mu}}_{\nu}$, into scalar-vector-tensor perturbations to give $\pis$, $\piv$ and $\pit$. In what follows we will concentrate on the scalar (density) perturbations, but the basic idea can also be applied -- more simply -- to the vector and tensor case, see for example \cite{PhysRevD.76.023005,Battye:2012eu, BattyePearson_connections}.

The perturbed fluid variables are constrained by the conservation equation, $\ep (\nabla_{\mu}U^{\mu\nu})=0$, which, in Fourier space for scalar perturbations in the synchronous gauge, are given by
\bse
\label{eq:sec:fluid-eqs-fourier}
\bea
\dot{\delta} &=& -(1+w)\bigg( - k^2\tis + \half\dot{h}\bigg) - 3 \hct w\Gamma,
\\
\tisdot &=& - \hct (1-3w)\tis - \frac{w}{1+w}\left(\delta +\Gamma - \frac{2}{3}\pis\right),
\eea
\ese
with the entropy perturbation
\bea
w\Gamma \defn \bigg( \frac{\delta P}{\delta\rho} - w\bigg)\delta  .
\eea
For simplicity we have set $\dot{w}=0$ and $\tis$ is defined from the velocity field via $\tis = i{\bf k}\cdot {\bf v}/k^2$.

Given these two conservation equations, knowledge of two of the fluid variables $\{\delta, \tis,\Gamma,\pis\}$ in terms of the other two and perturbed metric variables $\{h, \eta\}$ (and their time derivatives) would close the system of equations and allow observational quantities to be deduced in the linear regime. If $SO(3,1)$ reparameterization invariance is imposed on the underlying theory, the entropy $\Gamma$ and anisotropic stress $\pis$ are gauge invariant quantities and hence it seems sensible to aim to write $\Gamma\equiv\Gamma(\delta,\tis,h,\eta)$ and $\pis\equiv\pis(\delta,\tis,h,\eta)$ where it is implicit that we include derivatives of the quantities.

In \cite{Battye:2012eu, BattyePearson_connections} we worked  out the cases  where the dark sector only contains the metric, $\ld = \ld(g_{\mu\nu})$ and is time translation invariant (that is, $SO(0,1)$ invariant), and another case where the dark sector contains a scalar field $\phi$ and its Lorentz invariant kinetic scalar $\kin \defn - \tfrac{1}{2}\nabla^{\mu}\phi\nabla_{\mu}\phi$. In these two cases, the gauge invariant equations of state are respectively given by
\bse
\label{eq:sec:eos-simple}
\bea
\label{eq:sec:eos_lg}
\ld(g_{\mu\nu})  \Rightarrow   \left\{ \begin{array}{c} w\Gamma =  0,\\ w\pis = \tfrac{3}{2}(w-\qsubrm{c}{s}^2)\big[ \delta - 3 (1+w) \eta\big].\end{array}\right.
\\
\label{eq:sec:eos_ab}
\ld(\phi, \kin)  \Rightarrow   \left\{ \begin{array}{c} w\Gamma = (\qsubrm{c}{s}^2-w)\big[\delta - 3 \hct(1+w) \tis\big],\\ w\pis =0.\end{array}\right.
\eea
\ese
In each of these parameterizations there is only one free function of time: $\qsubrm{c}{s}^2(t)$ which can be interpreted as a speed of sound propagation.  In analogy to the case of $w$, we would envisage taking this to be constant until the data are sufficiently constraining to make a principal component analysis meaningful. (\ref{eq:sec:eos_lg}) describes perturbations of elastic dark energy \cite{PhysRevD.60.043505, PhysRevD.76.023005}. When $\qsubrm{c}{s}^2  \equiv 1$, (\ref{eq:sec:eos_ab}) describes minimally coupled quintessence perturbations and more general values correspond to $k$-essence models \cite{Weller:2003hw,PhysRevD.69.083503,Creminelli:2008wc}.

\textit{\textbf{Relation to the field content}} We now give an example field content to  show what types of theories yield equations of state.
We start from a field content given by
\bea
\label{eq:sec:fieldcontent-eos}
\ld = \ld(\phi, \partial_{\mu}\phi, \partial_{\mu}\partial_{\nu}\phi, g_{\mu\nu}, \partial_{\alpha}g_{\mu\nu}),
\eea
which is at most linear in $ \partial_{\alpha}g_{\mu\nu}$,
and $\ep U^{\mu\nu}$ can be computed from the quadratic Lagrangian for perturbations, $\sol$, and written as
\bea
\ep U^{\mu\nu} = \hat{\mathbb{Y}}^{\mu\nu}\lp\phi + \hat{\mathbb{W}}^{\mu\nu\alpha\beta} \lp g_{\alpha\beta}- \lied{\xi}U^{\mu\nu}.
\eea
We have written $\lp = \ep + \lied{\xi}$ (where $\lied{\xi}$ is the Lie derivative along $\xi^{\mu}$) so that the theory is built with the Stuckelberg fields $\xi^{\mu}$ which parameterize  violations of reparameterization invariance -- in everything that follows, the parameters in the theory are arranged so that the $\xi^{\mu}$-fields consistently decouple and the theory is manifestly $SO(3,1)$ reparameterization invariant.
We restrict ourselves (for calculability) to the theories which have
\bse
\label{eq:sec:deriv-expn}
\bea
\hat{\mathbb{Y}}^{\mu\nu}&=& \mathbb{A}^{\mu\nu}  + \mathbb{B}^{\alpha\mu\nu}\nabla_{\alpha}  + \mathbb{C}^{\alpha\beta\mu\nu} \nabla_{\alpha}\nabla_{\beta}  \nonumber\\
&&\qquad+ \mathbb{D}^{\rho\alpha\beta\mu\nu}\nabla_{\rho}\nabla_{\alpha}\nabla_{\beta} ,\\
\hat{\mathbb{W}}^{\mu\nu\alpha\beta} &=& \mathbb{E}^{\mu\nu\alpha\beta}  + \mathbb{F}^{\rho\mu\nu\alpha\beta}\nabla_{\rho} .
\eea
\ese 
The symmetries of the tensors $\{\mathbb{A}, \ldots, \mathbb{F}\}$ in (\ref{eq:sec:deriv-expn}) are inherited from the effective Lagrangian for perturbations \cite{Battye:2012eu}. One could include higher-derivative terms in the expansions which would encompass more theories.  We impose spatial isotropy on the background spacetime to allow use of the (3+1) decomposition to isolate all terms in the tensors in the expansions of $\hat{\mathbb{Y}}^{\mu\nu}, \hat{\mathbb{W}}^{\mu\nu\alpha\beta} $.  
 For theories which are (i) $SO(3,1)$ reparameterization invariant and  (ii) have second order field equations,  the perturbed fluid variables are constructed from the field variables according to
\bea
\label{eq:sec:fluid-vsrs-ppf}
\left( \begin{array}{c} \delta-A_{14}\dot{h} \\ \tis \\ \delta P  \end{array}\right) = \left( \begin{array}{ccc} \qsubrm{A}{11} & \qsubrm{A}{12} & 0 \\ \qsubrm{A}{21} & \qsubrm{A}{22} & 0  \\ \qsubrm{A}{31} & \qsubrm{A}{32} & \qsubrm{A}{33} \end{array}\right)\left( \begin{array}{c} \vphi \\ \dot{\vphi} \\ \ddot{\vphi} \end{array}\right) ,
\eea
and it turns out, in the case under consideration, that $\pis=0$ (and $\piv=\pit=0$);  in order to create a non-zero anisotropic stress, in the case of just scalar and tensor fields, it appears to be necessary to break at least part of the reparameterization invariance as is the case of elastic dark energy.
The  eight functions $\{A_{11},\ldots A_{33}\}$ are  only time dependent. It is possible to isolate   how the components of the tensors $\{\mathbb{A} , \ldots, \mathbb{F}\}$ combine to construct the $\qsubrm{A}{IJ}$  in (\ref{eq:sec:fluid-vsrs-ppf}); the exact form is very complicated, and we don't reproduce it here. As an example, one  finds that it is only the components of the tensor $\mathbb{F}$ which contribute to $A_{14}$.  If we now eliminate the internal degrees of freedom $\{\vphi, \dot{\vphi}, \ddot{\vphi}\}$, the entropy perturbation is of the form 
\bea
\label{eq:sec:entropy-notyetgaginv}
w\Gamma=B_1\delta+B_2\theta+B_3\dot{h}+B_4\ddot{h}
\eea
for some $\qsubrm{B}{I}$ which are functions of the $\qsubrm{A}{IJ}$. The dependancies of these functions are $B_1 = B_1(t), B_2 = B_2^{(0)} (t)+ k^2 B_2^{(1)}(t), B_3 = B_3(t), B_4 = B_4(t)$.

The functions $\qsubrm{A}{IJ}$ represent how the field content combines to construct the fluid variables; for this reason we call the $\qsubrm{A}{IJ}$ the \textit{activation functions}, or AFs for short, and we call the matrix $[\qsubrm{A}{IJ}]$ the \textit{activation matrix}. Specific theories can be used to prescribe the values of the AFs. Suppose that if for some reason one has $A_{33}=0$ then $\ddot{\vphi}$ is ``deactivated'' and does not influence the fluid variables.  The important take-away message  from (\ref{eq:sec:fluid-vsrs-ppf}) is to notice how the activation matrix prescribes which field variables are activated and thus appear in which fluid variable.

\textit{\textbf{Gauge invariant equations of state for perturbations}} Earlier we stated that the form of $w\Gamma$   (\ref{eq:sec:entropy-notyetgaginv}) must be gauge invariant. The most general gauge invariant form of the equation of state that yield second order field equations for all theories with field content (\ref{eq:sec:fieldcontent-eos}) and energy-momentum tensor (\ref{eq:sec:deriv-expn}) can be written as
\bea
\label{eq:Sec:eos_dspters-paper-result-tot}
\label{eq:Sec:eos_dspters-paper-result}
w\Gamma &=& (\alpha-w)\bigg[ \delta - 3\hct(1+w)\beta_1 \tis-\frac{3\hct(1+w)\beta_2}{2k^2-6(\dot{\hct}-\hct^2)}\dot{h}\nonumber\\
&&+\frac{3\hct(1+w)(1-\beta_1-\beta_2)}{6\ddot{\hct} + 6\hct^3 - 18\hct\dot{\hct}+2k^2\hct}\ddot{h}\bigg].
\eea
This is the \textit{equation of state for dark sector perturbations}, since it prescribes exactly how the details of a theory combine into terms which directly alter the fluid equations and cosmological observables. With (\ref{eq:Sec:eos_dspters-paper-result-tot}) and the fact that $\pis=0$, the    fluid equations (\ref{eq:sec:fluid-eqs-fourier}) close.  There are   $3$ dimensionless functions of scale and time $\mathcal{F}\defn\{\alpha, \beta_1, \beta_2\}$ which completely specify perturbations. The scale dependance of the functions $\mathcal{F}$   can be precisely isolated, reducing the freedom to a set of time-dependant functions only.

The functions  $\mathcal{F}$ can all be determined as functions of the AFs $\qsubrm{A}{IJ}$, which in turn can all be determined from components of the tensors (\ref{eq:sec:deriv-expn}) that were derived from a Lagrangian for perturbations. Particular theories can be used to  compute what these functions are. The functions  $\mathcal{F}$ are subject to stability requirements and thus cannot take arbitrary values. 

The number of free functions can be reduced by choosing restrictions on the theory space. There are a few obvious ways to do this, and each will end up simplifying the equation of state. For example,  if $A_{33}=0$   then $\ddot{\vphi}$ is removed from the ``active'' field content (\ref{eq:sec:fluid-vsrs-ppf}); this renders all $\qsubrm{B}{I}$ in (\ref{eq:sec:entropy-notyetgaginv}) scale independant and sets $\beta_2 = 1-\beta_1$ in the equation of state (\ref{eq:Sec:eos_dspters-paper-result-tot}). Another way to proceed could be to knock out   tensors from the expansion of $\ep {U^{\mu}}_{\nu}$. For instance,   when all components   $\mathbb{F}=0$, then $A_{14}=0$; this restriction also sets $\beta_2 = 1-\beta_1$. Note that if additionally $\beta_1=1$ then (\ref{eq:Sec:eos_dspters-paper-result}) becomes (\ref{eq:sec:eos_ab}).

\textit{\textbf{Example: Kinetic Gravity Braiding}}
For illustrative purposes, we     study a class of   examples that are included in the more general form of the AFs described earlier (and therefore of our equations of state for dark sector perturbations).  This can be thought of as being a ``truncated'' Horndeski theory, and is given by the   Kinetic Gravity Braiding \cite{Creminelli:2006xe, Creminelli:2008wc, Deffayet:2010qz, Pujolas:2011he, Kimura:2010di} (KGB) theory, whose Lagrangian is  
\bea
\label{eq:sec:lag-kgb}
\ld = \mathcal{A}(\phi, \kin)\square\phi+\mathcal{B}(\phi, \kin),
\eea
where  $\mathcal{A, B}$ are arbitrary functions of the scalar field $\phi$ and the kinetic scalar $\kin \defn - \half \nabla^{\mu}\phi\nabla_{\mu}\phi$. The field equations for the scalar field  are at most of second order. The energy-momentum tensor calculated from (\ref{eq:sec:lag-kgb}) is given by $U_{\mu\nu} = \ld_{,\kin} \nabla_{\mu}\phi\nabla_{\nu}\phi + 2 \nabla_{(\mu}\mathcal{A}\nabla_{\nu)}\phi + Pg_{\mu\nu}$,
where $P \defn \mathcal{B} - \nabla^{\mu}\phi\nabla_{\mu}\mathcal{A}$. On an isotropic background, the energy density $\rho$ and pressure $P$ are given by
\bse
\bea
\rho &=& - \mathcal{B} + 2 (\mathcal{A}_{,\phi} + \mathcal{B}_{,\kin}) - 2 \mathcal{A}_{,\kin}\kin\sqrt{2\kin}K\nonumber\\
&\defn&\rho(\phi, \kin,K),\\
P &=& \mathcal{B} + 2 \mathcal{A}_{,\phi} \kin + 2 \mathcal{A}_{,\kin}\kin\sqrt{2\kin}\mathcal{Y}\nonumber\\
&\defn& P(\phi, \kin,\mathcal{Y}),
\eea
\ese
 where $\mathcal{Y} \defn \dot{\kin}$, $K=3\hct$ and $\hct $ is the Hubble parameter. These can then be used to compute the following first variations,
\bea
\left( \begin{array}{c} \delta\rho-\tfrac{1}{2}\rho_{,K}\dot{h} \\ \tis \\ \delta P \end{array}\right) = \left( \begin{array}{ccc} \rho_{,\phi} & \rho_{,\kin} & 0 \\ b_1 & b_2 & 0  \\ P_{,\phi} & P_{,\kin}  & P_{,\mathcal{Y}} \end{array}\right)\left( \begin{array}{c} \vphi \\  \delta\kin \\ \delta\mathcal{Y} \end{array}\right) ,
\eea
where $b_1,b_2$ are functions of background field variables, given by
\bse
\bea
(\rho+P)b_1 &\defn& (\mathcal{B}_{,\kin} + 2 \mathcal{A}_{,\phi} - K\mathcal{A}_{,\kin}\sqrt{2\kin}) \sqrt{2\kin},\\
(\rho+P)b_2 &\defn& \mathcal{A}_{,\kin}\sqrt{2\kin}.
\eea
\ese
We can deduce that $B_2=B_2^{(0)}(t)+k^2B_2^{(1)}(t)$ with $B_1(t)$, $B_3(t)$ and $B_4(t)$ scale independent. From this we conclude that $\alpha\equiv\alpha(t)$, $\beta_1\equiv\beta_1^{(0)}(t)+k^2\beta_1^{(1)}(t)$ and $\beta_2\equiv[3(\dot\hct-\hct^2)+k^2]\beta_2^{(0)}(t)$, that is, there are just 4 time dependent functions which can be measured: $\alpha$, $\beta_1^{(0)}$, $\beta_1^{(1)}$, $\beta_2^{(0)}$ (NB the last two are not dimensionless). 
If the KGB theory is shift symmetric, that is, $\ld = \mathcal{A}(\kin)\square\phi+\mathcal{B}(\kin)$, then $\beta_1^{(0)}\equiv 0$ and therefore there are then just 3 time-dependent functions.

\textit{\textbf{Discussion}}
In this letter we outlined the philosophy of our approach, and we provided our main results with schematic derivations. These are the activation matrix (\ref{eq:sec:fluid-vsrs-ppf}) and the gauge invariant equations of state for dark sector perturbations  (\ref{eq:Sec:eos_dspters-paper-result}). The equations of state are a neat way to package parameterizations of perturbations in the dark sector and we advocate that those attempting to constrain dark energy and modified gravity theories of the kind described by ${\cal L}={\cal L}(\phi,\partial_\mu\phi,\partial_\mu\partial_\nu\phi,g_{\mu\nu},\partial_{\alpha}g_{\mu\nu})$
should do this via the parameters $\alpha$, $\beta_1$ and $\beta_2$ which can be related back to the fundamental nature of the theory. We have not included explicit expressions relating the different parts to our calculation. These will be presented in future work along with our analysis of the current observational constraints on the parameterization and the prospects for the future.

\textit{\textbf{Acknowledgements}} We have appreciated conversations with Tessa Baker, Alex Barreira, Pedro Ferreira, Adam Moss,   Ignacy Sawicki and Constantinos Skordis. JAP  is supported by the STFC Consolidated Grant ST/J000426/1. This research was supported in part by Perimeter Institute for Theoretical Physics. Research at Perimeter Institute is supported by the Government of Canada through Industry Canada and by the Province of Ontario through the Ministry of Economic Development and Innovation.

\end{document}